\documentclass[aps,prl,amsmath,twocolumn]{revtex4}
\renewcommand{\section}[1]{\paragraph{\textbf{#1}}}
\renewcommand{\section}[1]{\noindent\paragraph{\bf\emph{#1}}}

\usepackage{braket}
\newcommand{\rojo}[1]{{\color{black}#1}}
\newcommand{\gr}[1]{{\color{black}#1}}
\newcommand{\azul}[1]{{\color{black}#1}}

	\usepackage{amsmath}
	\usepackage{amsfonts,amsmath,amssymb}
  	\usepackage{amssymb}
  	\usepackage{float}
	\usepackage{makeidx}
	\usepackage{amsfonts}
	\usepackage[ansinew]{}
	\usepackage[usenames,dvipsnames]{pstricks}
	\usepackage{graphicx}

\usepackage{xcolor}

\newcommand{\be}{\begin{equation}}
\newcommand{\ee}{\end{equation}}
\newcommand{\bea}{\begin{eqnarray}}
\newcommand{\eea}{\end{eqnarray}}

\makeindex

\begin{document}


\title{Deep learning merger masses estimation from gravitational waves signals in the frequency domain}

\author{Juan Pablo Marulanda${}^{3}$}
\author{Camilo Santa${}^{2,3}$}
\author{Antonio Enea Romano${}^{1,2,3}$}

\affiliation{%
${}^{1}$Theoretical Physics Department, CERN,CH-1211 Geneva 23, Switzerland\\
${}^{2}$ICRANet, Piazza della Repubblica 10,I–65122 Pescara, Italy\\
${}^{3}$Instituto de Fisica,Universidad de Antioquia,A.A.1226, Medellin, Colombia
}%

\date{\today}

\begin{abstract}
Detection of gravitational waves (GW) from compact binary mergers provide a new window into multi-messenger astrophysics. The standard technique to determine the merger parameters is matched filtering, consisting in comparing the signal to a template bank. This approach can be time consuming and computationally expensive due to the large amount of experimental data which needs to be analyzed. 

In the attempt to find more efficient data analysis methods we develop a new frequency domain convolutional neural network  (FCNN) to predict the merger masses from the spectrogram of the detector signal, and compare it to time domain neural networks (TCNN). Since FCNNs are trained using spectrograms, the dimension  of the input is reduced as compared to TCNNs, implying a substantially lower number of model parameters, and consequently less over-fitting. The additional time required to compute the spectrogram is approximately compensated by the lower execution time of the FCNNs, due to the lower number of parameters.
In our analysis FCNNs show a slightly better performance on validation data and a substantially lower over-fit, as expected due to the lower number of parameters, providing a new promising approach to the analysis of GW detectors data, which could be further improved in the future by using more efficient and faster computations of the spectrogram.
\end{abstract}

\pacs{Valid PACS appear here}
\maketitle

\section{Introduction}

According to general relativity, gravitational waves  propagate at the speed of light and in the linear perturbative regime are produced by the second order time derivative of the quadrupole moment. The main sources of gravitational waves that can be detected at the Laser Interferometer Gravitational Wave Observatory (LIGO) and Virgo collaborations are the mergers of compact binary systems composed by black holes or neutron stars. 

Due to the no-hair theorem, the merger of two black holes has no electromagnetic counterpart \cite{Abbott:2019yzh} but binary star mergers can have electromagnetic signals, therefore opening a new window into multi-messenger astronomy. These systems are known as standard sirens because the gravitational wave signal provides information of the distance to the objects independent of the cosmic distance ladder and their electromagnetic counterparts provide information about their speed. Therefore, they can be used to measure the Hubble parameter \cite{Schutz:1986gp,Abbott:2019yzh} or to constrain alternative gravity theories with superluminal or subluminal GW  speeds \cite{Monitor:2017mdv}.

In order to achieve this goal, the detectors must have a strain sensitivity of the order of $10^{-21}/\sqrt{Hz}$ \cite{Martynov:2016fzi} and the standard data analysis approach consist in using matched filtering to compare the detector signal to a bank of gravitational wave templates in order to determine the merger parameters. Neural networks can be used to denoise the raw signal \cite{Shen:2017jkj,Wei:2019zlc} as a preprocessing step before matched filtering. This data analysis process must be repeated for every signal, which can be very time consuming and computationally expensive depending on the size of the template bank. Another approach has been developed \cite{George:2016hay,George:2017pmj,George:2017vlv,Rebei:2018lzh,Gabbard:2017lja,Fan:2018vgw,Gonzalez:2018mfo,Chua:2018woh,Nakano:2018vay} in which the time domain detector data is processed by a convolutional neural network to predict the  merger masses. In this paper we present the results of applying CNN to the frequency domain data, i.e. the Fourier transform of the time domain data, and call this neural network FCNN, to distinguish it from the CNN applied to time domain data, which we denote as TCNN.

The FCNN relies on the short-time Fourier transform to extract the frequency domain features needed to train the network. This approach reduces the dimensionality of the input, and the FCNN has around \rojo{50.000} parameters, compared with almost 550.000 of the TCNN. As a direct consequence  FCNN have better out of sample performance compared to TCNN, and tend to also have a lower  over-fit, due to the significantly lower model complexity.

\section{Training Data generation}

The training data is generated using the PyCBC package\footnote{Visit https://pycbc.org/ for more info.}, developed by the PyCBC Development Team and the LIGO / Virgo Collaborations.
	
\begin{figure}[h]
    \centering
    \includegraphics[scale=0.8]{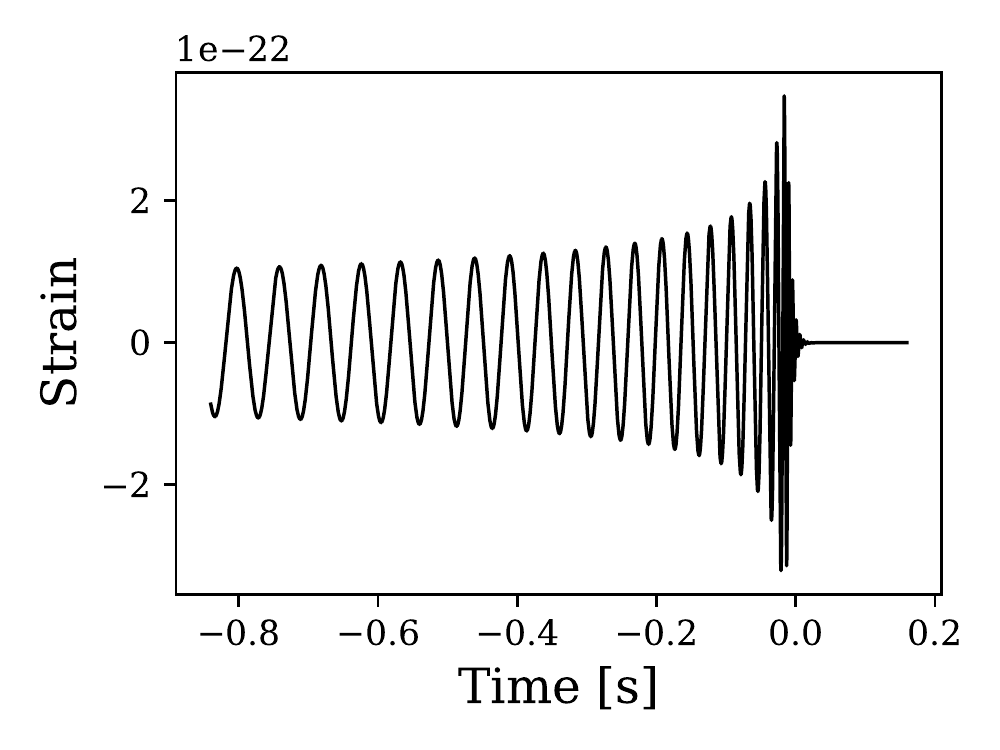}
    \caption{\rojo{Simulated strain of a black hole merger with 57 and 33 solar masses at a distance of 2000 MPc, sampled at 8192 Hz.}}
    \label{fig.strains}
\end{figure}

This library contains a method to generate the waveform corresponding to a GW event, and accepts as inputs several different parameters. In the waveform generation we assumed for simplicity the spins and orbital eccentricities to be zero, as  in \cite{George:2016hay}. Data with $\pi /2$ polarization was also generated in order to \azul{evaluate the robustness of the neural network to signals with different parameters}. The networks were trained to predict the two masses of the merger, while other parameters were kept fixed in the data generation.
We kept the default values for the other parameters of the waveform generator function except for 
the approximant, which is chosen to be the fourth version of Spin Effective One Body Numerical Relativity (SEOBNR)  due to its efficiency. \rojo{An example of a merger GW simulated signal is shown in fig.(\ref{fig.strains}).}

In order to train the networks with realistic data we add noise to the simulated signal. \rojo{ Similarly to \cite{George:2016hay}, in order to account for translations in the signal, the data was augmented by applying a random temporal shift in the interval [0,0.2].} 
We generated data with different signal-to-noise ratios (SNR), and 
colored noise was added according to the power spectral density (PSD) provided by LIGO. 

\begin{figure}[h]
    \centering
    \includegraphics[scale=0.8]{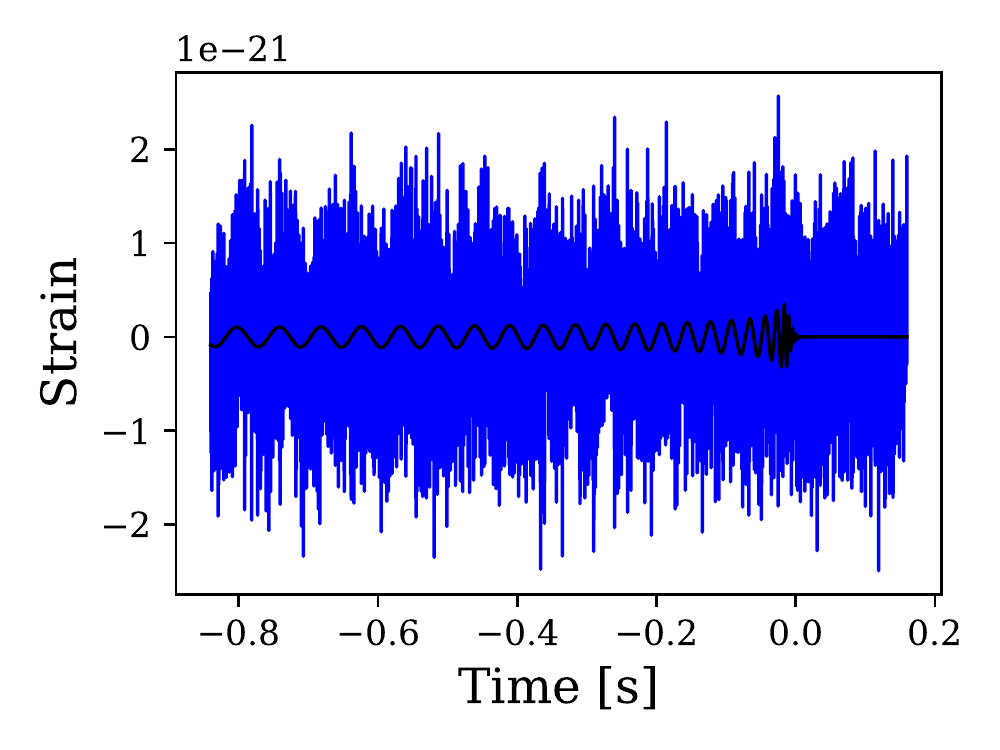}
    \caption{\rojo{Simulated strain superimposed with noise sampled from LIGO's aLIGOZeroDetHighPower PSD, with a SNR of 18.5.}}
    \label{fig.colored_strains}
\end{figure}

The matched-filter SNR between a template $h$ and a signal $s$ is defined by \cite{Raman_2018}:
\begin{equation}
    SNR=\frac{|\braket{s|h}(t)|^2}{\braket{h|h}},
\end{equation}
the bracket notation involves the following correlation:
\begin{equation}
    \braket{s|h}(t)=4\int_0^\infty \frac{\hat{s}(f)\hat{h}(f)}{S_n(f)} e^{2\pi i f t} df,
\end{equation}
where $\hat{s}$ and $\hat{h}$ are the Fourier transforms of the signal and template respectively and $S_n$ is the \azul{PSD} of the detector.
An example of the signal and the corresponding noised signal is shown in fig.(\ref{fig.colored_strains}).

We  train the FCNN  using spectrograms, which are two dimensional matrices whose  columns are related to the frequency power spectra of the strain ST at different times, according to

\be
SP_{\omega}=10*\log_{10}(|ST_{\omega}|^2) \, ,
\ee
where $SP_{\omega}$ is the spectrogram, and $ST_{\omega}$ is the Fourier transform of ST over different time intervals.

The spectrograms are obtained by performing a Fast Fourier Transform (FFT) using equally spaced time intervals, with  sampling frequency of \rojo{4096 Hz, windows of 128 elements, a zero-padding of 896 elements, and an overlap between windows of 64 elements.} 

\begin{figure}[h]
    \centering
    \includegraphics[scale=0.8]{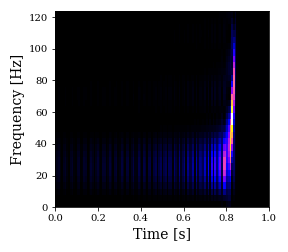}
    \caption{Spectrogram of the GW signal of a binary black hole merger with 33 and 57 solar masses.} 
    \label{fig.clean_freq}
\end{figure}

\begin{figure}[h]
    \centering
    \includegraphics[scale=0.7]{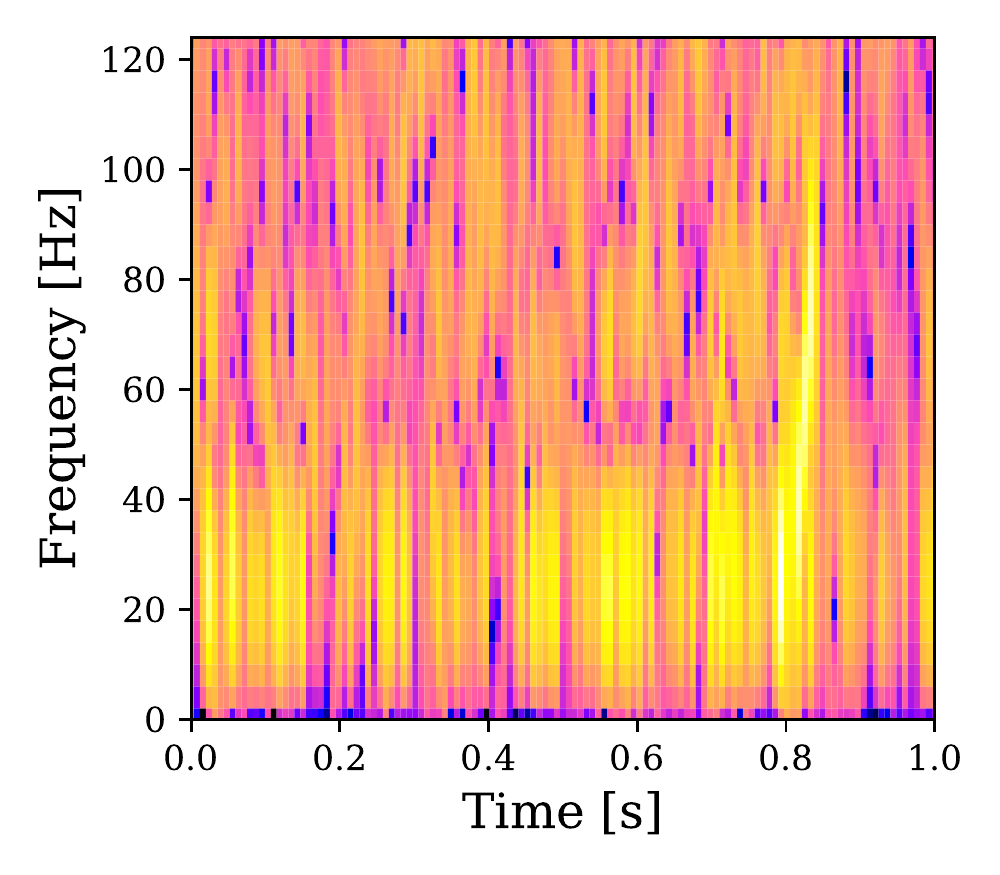}
    \caption{\rojo{Spectrogram of the noised GW signal of a binary black hole merger of 57 and 33 solar masses and SNR=18.5.}}
    \label{fig.colored_freq}
\end{figure}

\rojo{The spectrogram of a merger GW signal is shown in fig.(\ref{fig.clean_freq}), and the spectrogram of the corresponding noised signal is shown in fig.(\ref{fig.colored_freq}), where it can be seen that the merger signal is mainly noticeable at low frequency. As a consequence, for the purpose of training the FCNN, the spectrograms were cropped at 120 Hz.}


\section{CNN architectures}

The TCNN described in \cite{George:2016hay},  summarized in Table \ref{tab:archCNN}, was implemented as a benchmark to compare the performance of the FCNN. 

The FCNN, whose architecture is shown in Table \ref{tab:archNewCNN}, consists of three convolutional layers that perform 2D convolutions on the zero-padded signal, followed by a max pooling layer. The resulting output of the pooling layer is then flattened into a one dimensional vector of 1024 entries, which is fed into a two layer fully connected net, that predicts the two masses of the merger.

The FCNN has about 50.000 parameters, compared with almost 550.000 of the TCNN.  The smaller number of parameters \azul{has a regularization effect by reducing} the variance of the model, making it less prone to over-fitting as the number of degrees of freedom is greatly reduced. This is achieved because the spectrogram reduces the total number of input components, the number of convolutions is less than the TCNN and the two dimensional pooling operation reduces the number of components more than the one dimensional pooling. \azul{The latter greatly reduces the number of input components before the flatten layer and the number of parameters in the subsequent dense layers}.


\begin{table}[h]
	\centering
	\begin{tabular}{c | c | c}
		  & Layer & Size \\
		\hline
		   & Input & vector ($8192$)\\
		 1 & Convolution (ReLu) & matrix ($8192 \times 16$) \\ 
		 2 & Pooling & matrix ($2048 \times 16$) \\
		 3 & Convolution(ReLu) & matrix ($2048 \times 32$) \\
		 4 & Pooling & matrix ($512 \times 32$) \\
		 5 & Convolution (ReLu) & matrix ($512 \times 64$) \\
		 6 & Pooling & matrix ($128 \times 64$) \\
		 7 & Flatten & vector($8192$) \\
		 8 & Dense layer (ReLu) & vector ($64$) \\
		   & Output & vector ($2$)
	\end{tabular}
	\caption{Architecture of the TCNN used in \cite{George:2016hay}.}
	\label{tab:archCNN}
\end{table}

\begin{table}[h]
	\centering
	\begin{tabular}{c | c | c}
		  & Layer & Size \\
		\hline
		   & Input & matrix ($32 \times 127 \times 1$)\\
		 1 & Convolution (ReLu) & matrix ($32 \times 127 \times 16$) \\ 
		 2 & Convolution (ReLu) & matrix ($32 \times 127 \times 8$) \\
		 3 & Convolution (ReLu) & matrix ($32 \times 127 \times 4$) \\
		 4 & Pooling & matrix ($8 \times 32 \times 4$) \\
		 5 & Flatten & vector($1024$) \\
		 6 & Dense layer (ReLu) & vector ($32$) \\
		 7 & Dense layer (ReLu) & vector ($16$) \\
		   & Output & vector ($2$)
	\end{tabular}
	\caption{Architecture of the FCNN.}
	\label{tab:archNewCNN}
\end{table}


\rojo{In order to improve the performance of the models, the input data was normalized before training. The normalization that gave best results was the min-max scaling defined by}

\begin{equation}
    SP_{\omega \, norm} = \frac{SP_{\omega} - \text{min} (SP_{\omega})}{\text{max}(SP_{\omega}) - \text{min} (SP_{\omega})}
\end{equation}


\section{Over-fit}



When the training set error is very low,
due to a high number of parameters, there is a risk
of over-fitting, which manifests in a large difference between the training and validation errors. In fact, even
if the error of the model on the training set reaches
low values, it does not necessarily imply that its predictive ability, when applied on data different from the
training set, will be as good. In order to quantify the
difference between the training and the validation errors we define the following over-fitting estimator


\begin{equation}\label{meanover-fit}
    \text{O} = \left| \frac{\text{train error } - \text{ test error}}{\text{test error}} \right| \,.
\end{equation}

\azul{Low values of the over-fitting estimator correspond to a small relative difference between the  training and validation errors, implying  the model will have a performance on out-of-sample data similar to the one on training data.}
  

\section{Training Metric}




The purpose of this model is to predict the masses of the merger from the spectrogram of the gravitational wave's strain signal. We train the model by iteratively minimizing the mean absolute percentage error (MAPE) each epoch, defined as

\begin{equation}
    MAPE = \frac{1}{n}\sum_{i=1}^{n}\sum_{j=1}^{2}\left| \frac{\hat{M}_{ij}- M_{ij}}{M_{ij}} \right| \,
\end{equation}

\noindent where $n$ is the number of samples in each epoch, $\hat{M}_{i1}$ and $\hat{M}_{i2}$ are the masses of the merger predicted by the model, while $M_{i1}$ and $M_{i2}$ are the masses from the training set used in the simulation for the $i-th$ sample.

\section{Comparing FCNN to TCNN performance}

\begin{figure}[h]
    \centering
    \includegraphics[scale=0.8]{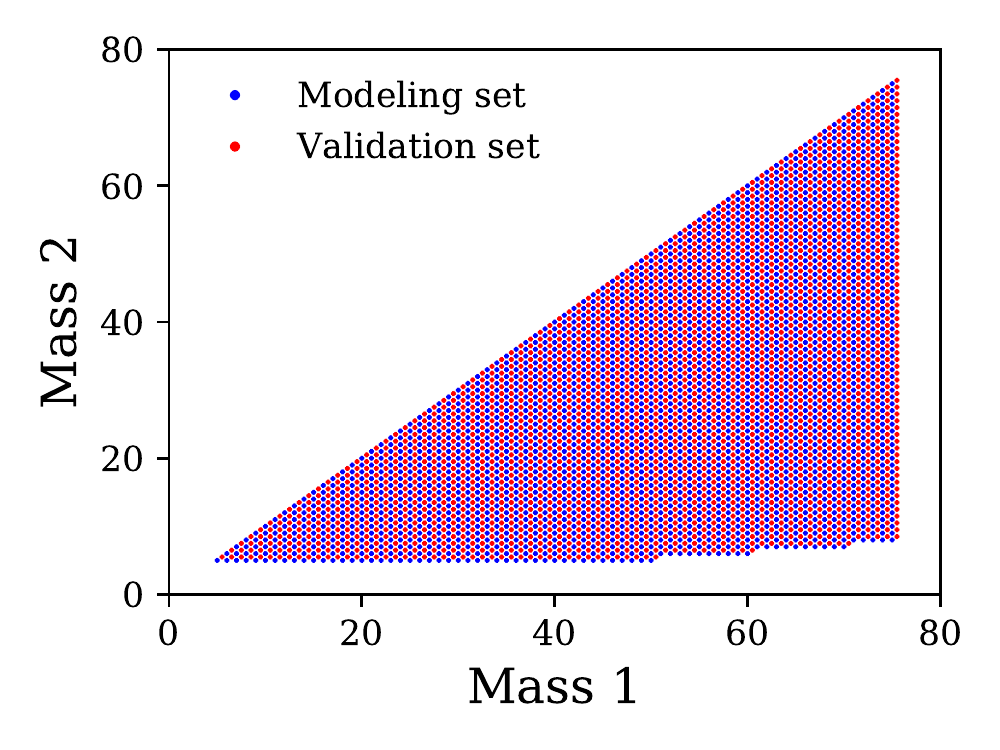}
    \caption{\rojo{Scatter plot of the mass pairs used to simulate the data. The modelling and validation data sets are chosen as in \cite{George:2016hay}.}}
    \label{fig.mass_dist}
\end{figure}

\rojo{
The merger GW data was simulated with a sampling rate of 8192 Hz, generated for mass values from $10M_{\odot}$ to  $75M_{\odot}$, with a mass ratio less than 10, and a distance of 2000 MPc, resulting in a total of 9346 mergers with SNRs in the  5 to 25 range. The data was split evenly between modeling and validation sets as shown in fig.(\ref{fig.mass_dist}). The modeling set was further split with a 70/30 ratio into training and development sets respectively, where the training set was used to optimize the model parameters, and the development set was used for hyperparameter tuning, specifically to find better performing  network architectures. The same procedure was applied to the two data sets simulated with polarization equal to 0 and $\pi/2$.}

\begin{figure}[h]
    \centering
    \includegraphics[scale=0.8]{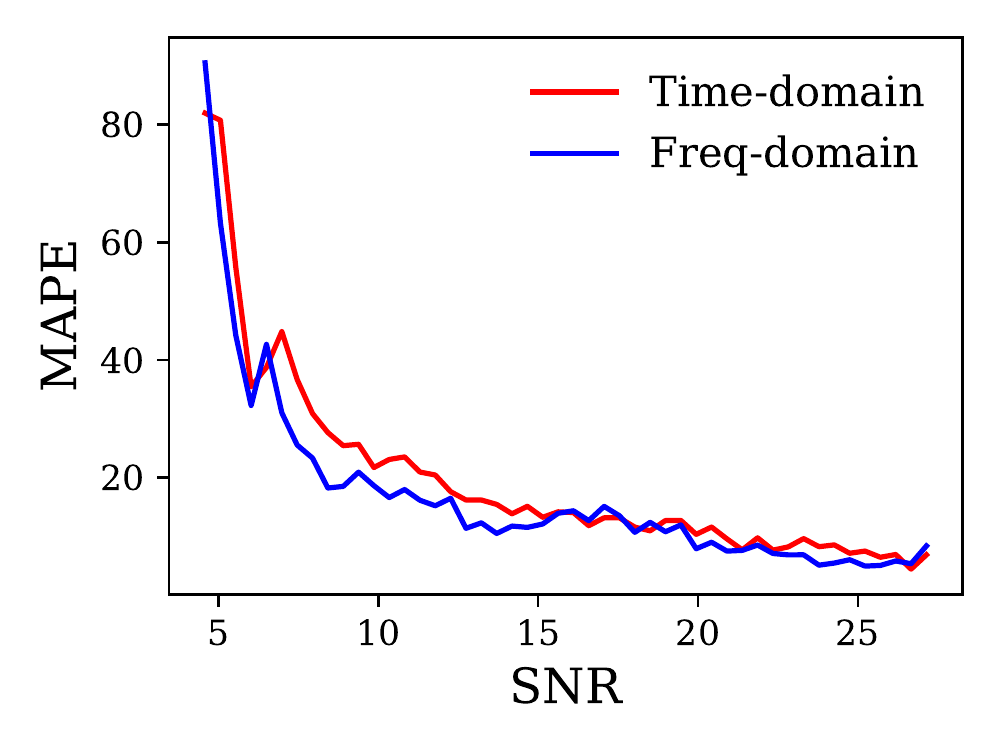}
    \caption{\rojo{The MAPE of the TCNN and FCNN models, evaluated  on the validation data set with a polarization equal to 0, is plotted as a function of the SNR.}}
    \label{fig.p0_val}
\end{figure}

The error of FCNN and TCNN models on the validation data over the range of SNRs is shown in fig.(\ref{fig.p0_val}) for data with polarization equal to 0.

\begin{figure}[h]
    \centering
    \includegraphics[scale=0.8]{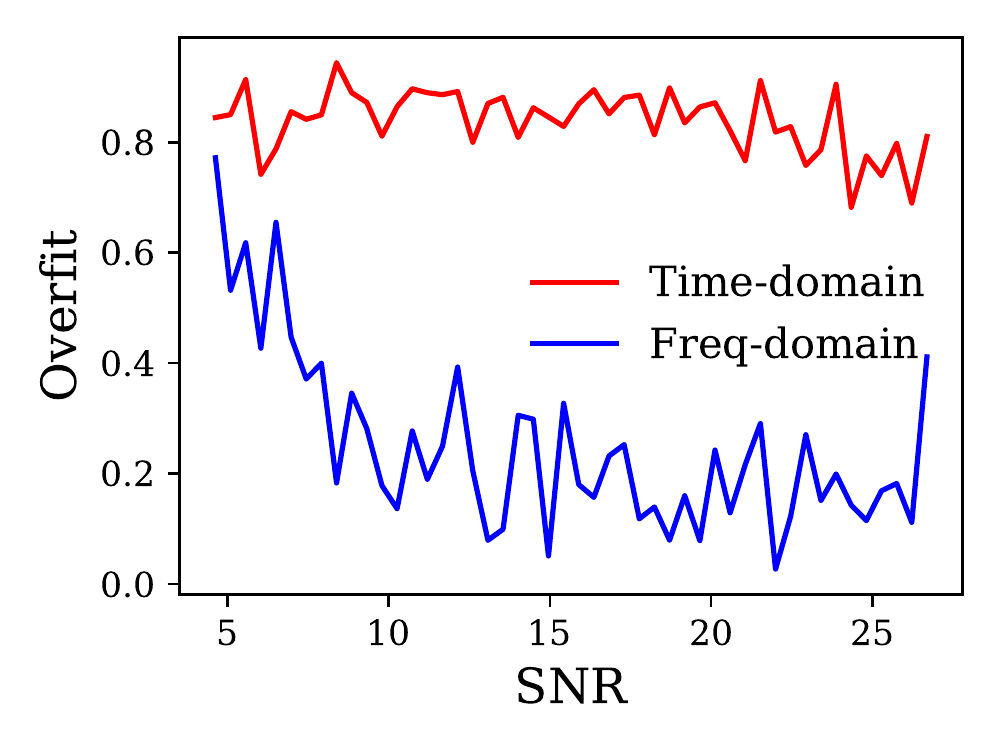}
    \caption{\rojo{The over-fit estimator, as defined in eq.(\ref{meanover-fit}), is plotted for TCNN and FCNN models as a function of the SNR, for data with polarization equal to 0.}}
    \label{fig.over_plot}
\end{figure}

In fig.(\ref{fig.over_plot}) the over-fit of the two models over the range of SNRs is shown, as defined in  eq.\eqref{meanover-fit}. As mentioned earlier, the FCNN has much fewer parameters than TCNN, and it is therefore  expected to have a lower over-fit than TCNN.


In order to test the robustness of the models under the change of other merger parameters we also created another set of training and validation data, using  a different polarization angle. Gravitational wave signals with a polarization angle of $\pi /2$ were simulated keeping all other parameters fixed. We used the same number of simulated samples, and the masses of the mergers ranged from $10M_{\odot}$ to  $75M_{\odot}$ with a mass ratio less than or equal to 10.

\begin{figure}[h!]
    \centering
    \includegraphics[scale=0.8]{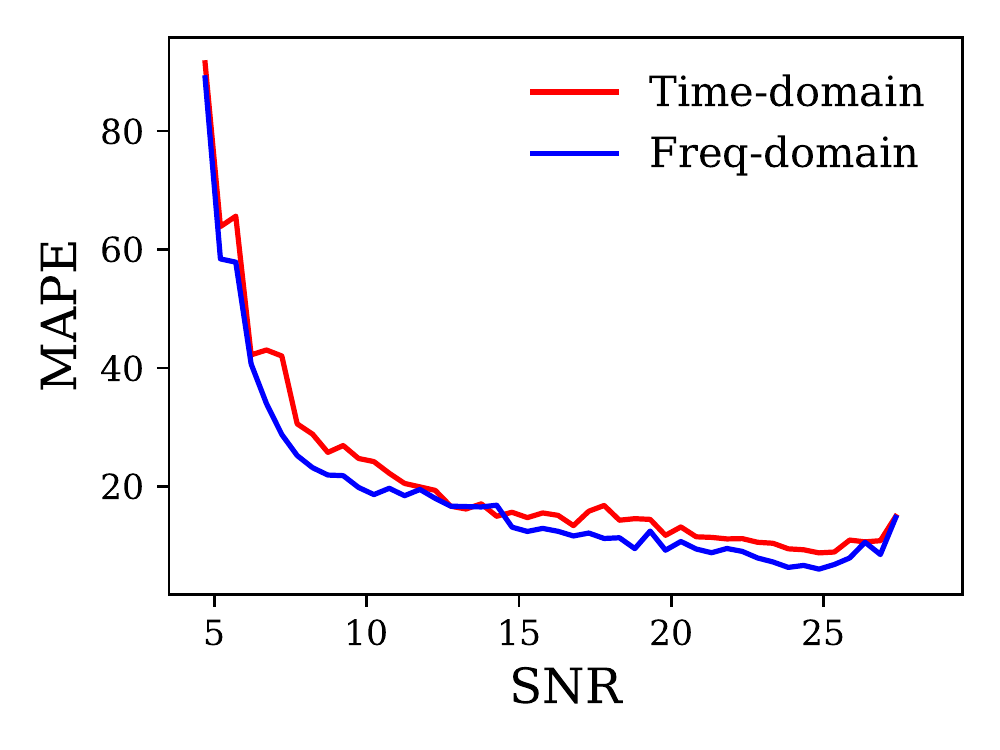}
    \caption{\rojo{The MAPE of the TCNN and FCNN models, evaluated  on the validation data set with a  polarization equal to $\pi/2$, is plotted as a function of the SNR.}}
    \label{fig.pi2_val}
\end{figure}
\begin{figure}[h!]
    \centering
    \includegraphics[scale=0.8]{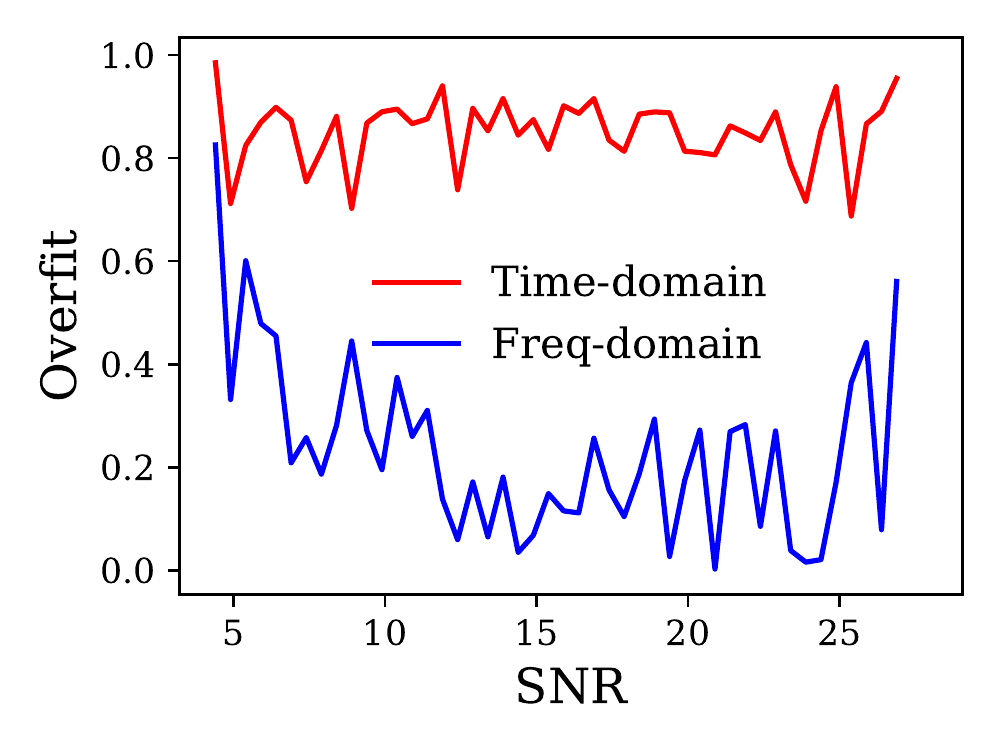}
    \caption{\rojo{The over-fit estimator, as defined in eq.(\ref{meanover-fit}), is plotted for TCNN and FCNN models as a function of the SNR, for data with polarization equal to $\pi/2$.}}
    \label{fig.over-fit_npol}
\end{figure}

The error of the TCNN and FCNN models on the validation set for data  with polarization equal to $\pi /2$ is shown in fig.(\ref{fig.pi2_val}). Likewise, the over-fit of TCNN models for this data is shown in fig.(\ref{fig.over-fit_npol}). The FCNN over-fit was lower than for TCNN,  suggesting that FCNN  generalizes better than TCNN also on signals from gravitational waves with different parameters.

\begin{figure}
    \includegraphics[scale=0.7]{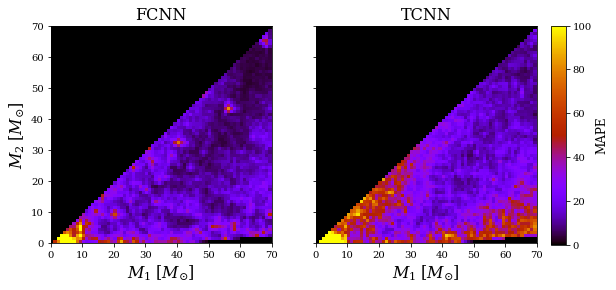}
    \caption{The MAPE of the prediction for out sample data with SNR=10 is plotted for different mass pairs.}
    \label{fig.mape_mass}
\end{figure}

\gr{As can be seen in fig.(\ref{fig.p0_val}) and fig.(\ref{fig.pi2_val}),  the performance of TCNN and FCNN is approximately the same, but for low SNRs the FCNN is slightly better. Nonetheless, from the over-fit plots it can be seen that, thanks to the reduced number of parameters,  FCNN has a better out-of-sample performance. A comparison of the MAPE of the mass predictions for out of sample data is shown in fig.(\ref{fig.mape_mass}).

The execution time of the FCNN is in general much lower than the TCNN, because the FCNNs have much less parameters.
If we add to this execution time the time necessary to compute the spectrogram using scipy.signal.spectrogram, we obtain a total computational time which is on average only about $6\%$ greater than that of a CNN working on the time domain data, but with a better MAPE and less over-fit due to the smaller number of parameters. Using more efficient implementations of the FFT to compute the spectrogram, and parallelizing it, could allow to reduce the FCNN pipeline execution time.
}

\onecolumngrid

\twocolumngrid

\section{Conclusions}

We have developed a new convolutional neural network, FCNN, to determine the  merger masses, trained on the spectrograms of simulated GW signals, and compared its performance with other CNN trained on time domain data (TCNN) \cite{George:2016hay}. 
The networks were trained for 1000 epochs using 4673 gravitational wave signals with a 70-30 train/development split, and the cost function which was minimized was the sum of mean absolute percentage errors between the masses and their predictions.
The FCNN was trained on spectrograms, allowing it to reduce the dimension of the input, resulting in a lower number of parameters in the final fully connected layers of the network, reducing its variance.

\gr{The execution time of the FCNN is in general much lower than the TCNN, because the FCNNs have much less parameters.
Adding the computational time  of the spectrogram using scipy.signal.spectrogram, we obtain a total time which is on average only about $6\%$ greater than that of a CNN working on the time domain data, but with a slightly better MAPE and substantially less over-fit, due to the smaller number of parameters.

In the future it could be interesting to use more efficient computations of the  spectrogram  using parallelization, in order to reduce the FCNN pipeline execution time.}
It will also be important to design  more complex FCNNs in order to predict additional parameters such as the spins and orbital eccentricities, or to apply it to data simulated for other detectors such a the Laser Interferometer Space Antenna.

\section{Acknowledgements}
We thank the anonymous Referee for helpful suggestions to improve the manuscript. 
This work was supported by the UDEA Dedicacion exclusiva and Sostenibilidad programs
and the CODI projects 2015-4044 and 2016-10945.

\bibliography{Bibliography}

\begin{thebibliography}{10}

\bibitem{Abbott:2019yzh}
LIGO Scientific, Virgo, B.~P. Abbott {\em et~al.},
\newblock (2019), arXiv:1908.06060.

\bibitem{Schutz:1986gp}
B.~F. Schutz,
\newblock Nature {\bf 323}, 310 (1986).

\bibitem{Monitor:2017mdv}
LIGO Scientific, Virgo, Fermi-GBM, INTEGRAL, B.~P. Abbott {\em et~al.},
\newblock Astrophys. J. {\bf 848}, L13 (2017), arXiv:1710.05834.

\bibitem{Martynov:2016fzi}
B.~P. Abbott {\em et~al.},
\newblock Phys. Rev. {\bf D93}, 112004 (2016), arXiv:1604.00439,
\newblock [Addendum: Phys. Rev.D97,no.5,059901(2018)].

\bibitem{Shen:2017jkj}
H.~Shen, D.~George, E.~A. Huerta, and Z.~Zhao,
\newblock (2017), arXiv:1711.09919.

\bibitem{Wei:2019zlc}
W.~Wei and E.~A. Huerta,
\newblock Phys. Lett. {\bf B800}, 135081 (2020), arXiv:1901.00869.

\bibitem{George:2016hay}
D.~George and E.~A. Huerta,
\newblock Phys. Rev. {\bf D97}, 044039 (2018), arXiv:1701.00008.

\bibitem{George:2017pmj}
D.~George and E.~A. Huerta,
\newblock Phys. Lett. {\bf B778}, 64 (2018), arXiv:1711.03121.

\bibitem{George:2017vlv}
D.~George and E.~A. Huerta,
\newblock {Deep Learning for Real-time Gravitational Wave Detection and
  Parameter Estimation with LIGO Data},
\newblock in {\em {NiPS Summer School 2017 Gubbio, Perugia, Italy, June 30-July
  3, 2017}}, 2017, arXiv:1711.07966.

\bibitem{Rebei:2018lzh}
A.~Rebei {\em et~al.},
\newblock Phys. Rev. {\bf D100}, 044025 (2019), arXiv:1807.09787.

\bibitem{Gabbard:2017lja}
H.~Gabbard, M.~Williams, F.~Hayes, and C.~Messenger,
\newblock Phys. Rev. Lett. {\bf 120}, 141103 (2018), arXiv:1712.06041.

\bibitem{Fan:2018vgw}
X.~Fan, J.~Li, X.~Li, Y.~Zhong, and J.~Cao,
\newblock Sci. China Phys. Mech. Astron. {\bf 62}, 969512 (2019),
  arXiv:1811.01380.

\bibitem{Gonzalez:2018mfo}
J.~A. Gonzalez and F.~S. Guzman,
\newblock Phys. Rev. {\bf D97}, 063001 (2018), arXiv:1803.06060.

\bibitem{Chua:2018woh}
A.~J.~K. Chua, C.~R. Galley, and M.~Vallisneri,
\newblock Phys. Rev. Lett. {\bf 122}, 211101 (2019), arXiv:1811.05491.

\bibitem{Nakano:2018vay}
H.~Nakano {\em et~al.},
\newblock Phys. Rev. {\bf D99}, 124032 (2019), arXiv:1811.06443.

\bibitem{Raman_2018}
A.~Raman,
\newblock 2018 IEEE Global Conference on Signal and Information Processing
  (GlobalSIP)  (2018).

\end{thebibliography}
\bibliographystyle{h-physrev4}

\end{document}